\documentclass[12pt]{article}

\usepackage{graphicx,subfigure}

\def\arcsec   {\hbox{$^{\prime\prime}$}}

\def\arcpt    {${{\lower3pt\hbox{$^{\prime\prime}$}}\atop{\raise4pt\hbox{.}}}$}

\def\To       {\ifmmode {\Theta_0}\else {$\Theta_0$}\fi}
\def\Ro       {\ifmmode {R_0}\else {$R_0$}\fi}
\def\vsun     {\ifmmode {V_\odot}\else {$V_\odot$}\fi}
\def\kms      {\ifmmode {{\rm km s}^{-1}}\else {km s$^{-1}$}\fi}
\def\msun     {\ifmmode {{\rm M}_\odot}\else  {${\rm M}_\odot$}\fi}
\def\Msun     {\ifmmode {{\rm M}_\odot}\else  {${\rm M}_\odot$}\fi}
\def\lsun     {\ifmmode {{\rm L}_\odot}\else  {${\rm L}_\odot$}\fi}
\def\vmax     {\ifmmode {V_{max}}\else {$V_{max}$}\fi}
\def\deg      {\ifmmode {\rlap.}^{\circ}\!\else ${\rlap.}^{\circ}\!$\fi}

\begin{document}
%%%%%%%%%%%%%%%%%%%%%%%%%%%%%%%%%%%%%%%%%%%%%%%%%%%%%%%%%%%%%%%%%%%%%%

\centerline{}
\vskip50pt
\huge
\centerline{\bf Ground-Based Astrometry 2010-2020}
\vskip50pt

\large
\centerline{\bf Todd J. Henry}
\vskip05pt
\centerline{\it Georgia State University, Atlanta, GA}
\vskip00pt
\centerline{\it 404-413-6054 / thenry@chara.gsu.edu}
\vskip15pt
\centerline{\bf David G. Monet, Paul D. Shankland}
\vskip05pt
\centerline{\it U.S.~Naval Observatory, Flagstaff, AZ}
\vskip15pt
\centerline{\bf Mark J. Reid}
\vskip05pt
\centerline{\it Harvard-Smithsonian Center for Astrophysics, Cambridge, MA}
\vskip15pt
\centerline{\bf William van Altena}
\vskip05pt
\centerline{\it Yale University, New Haven, CT}
\vskip15pt
\centerline{\bf Norbert Zacharias}
\vskip05pt
\centerline{\it U.S.~Naval Observatory, Washington, DC}

\normalsize
\vskip50pt
\centerline{\it also read and endorsed by:}
\vskip15pt

G.~Fritz Benedict, University of Texas, Austin, TX

Dana Casetti-Dinescu, Yale University, New Haven, CT
 
Ralph A.~Gaume, U.S.~Naval Observatory, Washington, DC

Terry Girard, Yale University, New Haven, CT

Wei-Chun Jao, Georgia State University, Atlanta, GA

Steven R. Majewski, University of Virginia, Charlottesville, VA

Brian D.~Mason, U.S.~Naval Observatory, Washington, DC

Richard J.~Patterson, University of Virginia, Charlottesville, VA

Imants Platais, Johns Hopkins University, Baltimore, MD

John P. Subasavage, Georgia State University, Atlanta, GA

Stephen C.~Unwin, Jet Propulsion Laboratory, Pasadena, CA

\vfil\eject

%%%%%%%%%%%%%%%%%%%%%%%%%%%%%%%%%%%%%%%%%%%%%%%%%%%%%%%%%%%%%%%%%%%%%%
\Large
\centerline{\bf Abstract}
\vspace{20pt}
\normalsize
%%%%%%%%%%%%%%%%%%%%%%%%%%%%%%%%%%%%%%%%%%%%%%%%%%%%%%%%%%%%%%%%%%%%%%

We discuss the opportunities for astronomical discovery via
ground-based astrometry carried out in the United States during the
2010-2020 decade.  We describe imminent scientific breakthroughs that
can be achieved at both classic astrometric scales --- narrow angle
astrometry done by individual groups and large A*$\Omega$ astrometry
carried out by consortia.  The two most compelling questions to be
addressed are (1) {\it What is the composition of the stellar and
substellar population near the Sun?} and (2) {\it What are the shape,
size, and mass of the Milky Way?}  We provide a short list of five
recommendations that we believe will allow us to take best advantage
of the intellectual and financial investments made for what some have
called ``The Decade of Astrometry.''  The most important
recommendation is to provide the educational foundation required so
that a new generation of astrometrists can make best use of the rich
datasets that will arrive in the coming decade.

%%%%%%%%%%%%%%%%%%%%%%%%%%%%%%%%%%%%%%%%%%%%%%%%%%%%%%%%%%%%%%%%%%%%%%
\vspace{-10pt}
\section{Introduction}
%%%%%%%%%%%%%%%%%%%%%%%%%%%%%%%%%%%%%%%%%%%%%%%%%%%%%%%%%%%%%%%%%%%%%%

Without a map, astronomers are lost.

The astronomical endeavor to understand the positions of objects in
the sky, or astrometry, is the foundation of our comprehension of the
Universe.  Astrometry provides both the (X,Y) positions on the sky
necessary for coordinates, as well as the all-important third
dimension, distance, that astronomers spend so much of their time
trying to measure.  One of the basic, yet most important, measurements
in astronomy is the trigonometric parallax, which forms the
fundamental rung in the astronomical distance ladder.

Advances in astrometry have always been driven by technology,
advancing from the human eye to modern telescopes and space
satellites.  On the ground, CCD mosaic cameras on medium-sized
telescopes are now offering enormous amounts of data with
milliarcsecond (mas) relative positions, and radio interferometers
have already leapt into the microarcsecond ($\mu$as) regime with
recent results from the VERA radio interferometric array (Honma et
al.~2007) and the VLBA (Reid et al.~2008).  In space, we are poised to
reach such precision on a grand scale in optical astrometry with
missions such as {\it J-MAPS}, {\it Gaia}, and {\it SIM}.  Such
precision makes it possible to do research at the forefront of
astronomy using astrometric techniques in a way never before possible.

Our goal here is to anticipate the forthcoming breakthroughs in
astronomy given sufficient support for astrometric programs in the
coming decade.  We focus specifically on major questions that can be
solved by measuring parallaxes and proper motions, while leaving
detailed discussions of binary stars and exoplanet searches to other
white paper efforts.  Throughout, a central theme should be remembered
--- the next decade will deliver more astrometric data with greater
accuracy to more users than the sum of all previous astrometric
endeavors.  Surprises are guaranteed.

%%%%%%%%%%%%%%%%%%%%%%%%%%%%%%%%%%%%%%%%%%%%%%%%%%%%%%%%%%%%%%%%%%%%%%
\vspace{-10pt}
\section{A Note about Space-Based Missions}
%%%%%%%%%%%%%%%%%%%%%%%%%%%%%%%%%%%%%%%%%%%%%%%%%%%%%%%%%%%%%%%%%%%%%%

The Decadal Survey Committee will hear separately from those involved
in space-based astrometric efforts, so we only briefly mention them
here for context.  {\it Hipparcos} (Perryman 2008) revolutionized
astrometry a decade ago by providing positions and parallaxes for
fairly bright stars to $\sim$1.0 mas, and with refinement, the
accuracy for some stars is now $\sim$0.5 mas (van Leeuwen 2007).  {\it
HST's} Fine Guidance Sensors make it possible to obtain relative
parallaxes to 0.2 mas accuracy for dozens of fainter stars of special
astrophysical interest (e.g., Benedict et al.~2008), and to help
calibrate the Galactic Cepheid distance scale.

In astronomy, one can always play the waiting game until something
better comes along.  There are at least four significant space-based
efforts that may be launched in the next ten years, each of which has
its niche: ESA's {\it Gaia}, scheduled to launch in 2011 with final
results due in 2020, will increase distances accurate to 1\% from the
10$^5$ stars measured by {\it Hipparcos} to 10$^7$ stars with
magnitudes 6--15, with upper limits on parallaxes for perhaps 10$^9$
stars.  USNO's {\it J-MAPS}, slated for a 2012 launch, will measure
10$^7$ stellar parallaxes to 1 mas with magnitudes 0--14 (results due
in 2016), and creates a 20-year proper motion baseline for bright
stars when linked to {\it Hipparcos} (bright stars will saturate in
{\it Gaia}).  Japan's planned infrared {\it JASMINE} spacecraft
(launch date not set) targets 10$^8$ stars in the Galactic bulge to
{\it z} $\sim$ 14.  The future of NASA's {\it SIM} remains uncertain,
but it could observe 10$^4$ stars with precisions at the $\mu$as level
and would reach to magnitude 20.  Even if all four are successful, it
must be stressed that not all high-accuracy astrometry will be done
with space satellites --- narrow angle astrometry in the next few
years will provide scientific breakthroughs that will merge into the
large A*$\Omega$ surveys that will soon follow.

%%%%%%%%%%%%%%%%%%%%%%%%%%%%%%%%%%%%%%%%%%%%%%%%%%%%%%%%%%%%%%%%%%%%%%
\vspace{-10pt}
\section{Narrow Angle Science: The Solar Neighborhood}
%%%%%%%%%%%%%%%%%%%%%%%%%%%%%%%%%%%%%%%%%%%%%%%%%%%%%%%%%%%%%%%%%%%%%%

The long history of ground-based astrometric work dates back to the
efforts of Hipparchus, Brahe, Galileo, and Kepler (among others), with
a specific breakthrough in 1838 by Friedrich Bessell, who measured the
first stellar parallax, for the nearby binary 61 Cygni.  The latest
ground-based parallax compendium is the General Catalog of
Trigonometric Parallaxes (van Altena et al.~1995), which includes
parallaxes for 8112 stars measured to a few mas.  Since then, roughly
700 total parallaxes have been published, typically accurate to
$\sim$1 mas (see {\it www.RECONS.org}).  The total number of targets
among current parallax programs is $\sim$1000.  The target lists come
from classic work such as Luyten's (1979), to more recent all-sky
surveys such as 2MASS, DENIS, and SDSS.  Follow-up narrow angle
investigations pinpoint the most intriguing sources, often at the
faint limits of such surveys, from which the science is squeezed out
when precise astrometric techniques are applied.  We list here some of
the most important questions that can be answered during the next
decade by targeted optical, infrared, and radio astrometry of
individual sources.  This work forms both the scientific and
educational foundations for our fundamental, and important, transition
to the large A*$\Omega$ projects that will expand each sample markedly
during the next decade.

$\bullet$ {\it What is the stellar population near the Sun?}  The
number of stellar systems known within 10 pc has increased by 18\%
since 2000 (Henry et al.~2006; {\it www.RECONS.org}), thereby tipping
the stellar mass function significantly to the smallest stars.  A more
accurate local census can be used to challenge star formation theory,
provide a benchmark for missing mass evaluations, and provide key
targets for major planet hunting missions.  A complete understanding
of the multiple star population will be forthcoming via detailed
astrometric studies at all separation scales larger than $\sim$10 mas.

$\bullet$ {\it What are the natures and number fractions of various
types of brown dwarfs?}  Great advances have been made in
understanding a few dozen substellar objects because parallaxes are
now available (Dahn et al.~2002; Vrba et al.~2004), but our knowledge
of their true space density is rudimentary at best.  High accuracy
astrometry at infrared wavelengths in particular is a nascent field,
and breakthroughs in brown dwarf research are imminent as large sky
surveys come online and push to fainter magnitudes.  By 2020, we will
know the relative populations of stars and brown dwarfs.

$\bullet$ {\it How many white dwarfs are there in the Galaxy?}  White
dwarfs are important chronometers for age-dating Galactic components,
but their number densities in the disk and halo are poorly known.
Even within 25 pc the number of known white dwarfs has climbed 25\%
since 2000 (e.g., Subasavage et al.~2009).  In addition to affecting
our understanding of the initial mass function, these are the objects
that will be used for severe tests of degenerate matter theory, as
they provide the highest fidelity spectra and will provide the most
accurate masses.

$\bullet$ {\it How much mixing of ancient subdwarfs is there in the
Sun's vicinity?}  Subdwarfs are outnumbered 100:1 by their main
sequence counterparts and are consequently poorly understood.  The
first comprehensive sample of subdwarfs within 60 pc is being
constructed now (Jao et al.~2009).  Continued identification of
subdwarfs via large-scale astrometric surveys with followup (or
concurrent) parallax efforts will lead to an enhanced understanding of
the complex age and metallicity mixing of the Galaxy's components.

$\bullet$ {\it What types of unseen companions are found orbiting
stars?}  One of the significant advantages of ground-based work is the
duration of the astrometric series.  The large surveys planned on the
ground and in space typically have five-year horizons, whereas some
ground-based surveys span a decade already (RECONS, USNO).  Such
surveys are exploring new territory for the types of unseen companions
sampled, including brown dwarfs and exoplanets with orbital periods
longer than Jupiter's.  In the foreseeable future, only ground-based
astrometry programs can provide multi-decade coverage.

$\bullet$ {\it What can stellar clusters tell us about stellar
evolution and populations?}  Star clusters provide us with
laboratories to study the formation and evolution of stars, with the
only major variable left being the stellar mass.  Current proper
motion measurements now allow us to identify cluster members to
magnitude 21, thereby permitting us to reach nearly the end of the
stellar main sequence out to 100 pc in various cluster environments
(Platais et al.~2003; Vieira et al.~2005).  High-precision astrometry
will also allow us to pin down the mass-luminosity relations in
fundamental clusters.

%%%%%%%%%%%%%%%%%%%%%%%%%%%%%%%%%%%%%%%%%%%%%%%%%%%%%%%%%%%%%%%%%%%%%%
\vspace{-10pt}
\section{Narrow Angle Science: The Milky Way and Beyond}
%%%%%%%%%%%%%%%%%%%%%%%%%%%%%%%%%%%%%%%%%%%%%%%%%%%%%%%%%%%%%%%%%%%%%%

Beyond the immediate solar neighborhood lurk the rare members of the
Milky Way and fundamental stellar clusters such as Orion, for which we
now have a distance known to better than 2\% (Menten et al.~2007).  We
are only just beginning to measure accurate distances to such sources,
and even today we only have an educated guess about the number of
spiral arms in our own Milky Way, or whether it is a barred Sb or Sc
galaxy.  Recent advances have been made at optical wavelengths with
improved detectors and observing techniques to achieve sub-mas
parallax errors, while at radio wavelengths, high-intensity masers can
be targeted interferometrically to provide parallaxes with errors as
small as 10 $\mu$as.  These capabilities open up new territory for
exploration through the Milky Way and into the hearts of nearby
galaxies.  A few parts of the territory that will expand in the next
decade are highlighted here.

$\bullet$ {\it What are the characteristics of rare members of the
Galaxy, such as cataclysmic variables and the central stars of
planetary nebulae?}  Concerted efforts on CVs at optical wavelengths
have yielded reliable distances for many, giving empirical heft to
estimates of luminosities, white dwarf masses, and (ultimately) space
densities (Thorstensen et al.~2008).  Because of ground-based parallax
efforts on planetary nebulae central stars, we now know their
luminosities and how they bridge the gap between massive stars and
white dwarfs (Harris et al.~2007).  Anticipated improvements in
parallax measurements will expand these samples by a factor of ten (or
more) in the coming decade.

$\bullet$ {\it What is the distribution of luminous and dark matter in
the Galactic disk?}  Different surveys give conflicting results,
although there may be some convergence towards the amount of luminous
matter being substantial and the amount of dark matter being small
(Girard et al.~2006).  To investigate, a survey with the WIYN One
Degree Imager covering 100 square degrees would provide parallaxes
with an accuracy of $\sim$0.5 mas to magnitude 21.  The derived robust
stellar and brown dwarf samples out to 150 pc will provide reality
checks for the large A*$\Omega$ surveys discussed below.

$\bullet$ {\it What are the shape, size, and mass of the Milky Way?}
Recent improvements in radio astrometry with the VLBA have yielded
parallaxes and proper motions to star forming regions across a
significant portion of the Milky Way with accuracies of $\sim$10
$\mu$as and a $\sim$1 km/s, respectively (Reid et al.~2008).  In the
next decade or so, we could measure the position to every high mass
star forming region in the Galaxy and provide a three-dimensional map,
complete with three-dimensional velocity vectors.  This would yield
the distance to the Galactic center (\Ro), the rotation speed of the
LSR (\To), the form of the rotation curve, and the kinematic effects
of spiral structure.  In conjunction with the
astrometrically-determined proper motions of Milky Way satellite
galaxies and improved numerical simulations, we will learn the size
and mass of the disk, bulge, and dark matter halo of the Milky Way.

$\bullet$ {\it How many dwarf galaxies are orbiting the Milky Way and
what are their orbits?}  Lambda cold-dark matter (LCDM) cosmological
models predict debris left from several hundred merging dwarf galaxies
within one kiloparsec of the Sun (Helmi \& White 1999).  To date, only
a handful of possible candidates have been identified, which casts
doubt on the validity of the model.  Alternatively, we may not yet
have found these faint objects.  The proper motions from deep, wide
surveys would reveal dozens of streams within a few kpc, if they
exist, thereby stress-testing the LCDM model.  Orbits such as those
measured by Dinescu et al.~(2005) for the Sgr dwarf are key to testing
the LCDM model on small scales where it is most challenged.  Such
orbits allow us to understand whether these satellites are of tidal
and/or cosmological origin (better defining the missing satellite
problem) and to estimate their masses by properly quantifying the
input from tides.

$\bullet$ {\it What is the proper motion of the Andromeda Galaxy?}
Andromeda's three-dimensional motion is key to understanding the
history and fate of the Local Group.  Andromeda's proper motion can be
measured via radio astrometry in the coming decade, and when combined
with available velocity information, its derived three-dimensional
motion will allow us to understand the delicate (or not) dance of
Andromeda and the Milky Way.

%%%%%%%%%%%%%%%%%%%%%%%%%%%%%%%%%%%%%%%%%%%%%%%%%%%%%%%%%%%%%%%%%%%%%%
\vspace{-10pt}
\section{Large A*$\Omega$ Surveys: Rich Data Mines}
%%%%%%%%%%%%%%%%%%%%%%%%%%%%%%%%%%%%%%%%%%%%%%%%%%%%%%%%%%%%%%%%%%%%%%

The grand ground-based sky surveys completed using the Palomar, UK,
and ESO Schmidt telescopes, and the subsequent digitizing of thousands
of photographic plates (USNO's PMM, U.~Edinburgh's SuperCOSMOS, and
STScI's DSS), revolutionized the way astronomers work by providing
huge datasets of astrometric and photometric data.  More recently,
SDSS, 2MASS, and DENIS have provided significant datasets using modern
CCD and infrared array detectors.  The next generation of ground-based
sky surveys is imminent, and these efforts will produce substantial
science before, in parallel, and beyond (in both time and phase space)
what space missions promise.  All of the previously mentioned science
avenues that profit from $\sim$1 mas parallaxes and proper motions
(and high quality photometric data) will benefit from these efforts,
with target lists that expand up to 10$^{10}$ objects.  Here we
describe four of the most comprehensive astrometric surveys, and then
concentrate on one application --- studies of the very nearby stars.

The key technologies that make the large etendue (A*$\Omega$) systems
possible are (a) large mosaics of CCDs, (b) the reduction in physical
size and power dissipation for the CCD electronics, (c) telescope
optics with exquisite image quality over a large fields of view, (d)
computer hardware and software that enable the scientific processing
of many terabytes of data/night, (e) databases that can store and
query these large data volumes, and (f) high-bandwidth communication
networks to distribute the results.  We are now experiencing nothing
less than a ``perfect storm'' of the necessary technologies.  Each of
the surveys discussed here can provide parallaxes (or upper limits)
for every star in the sky surveyed to their limiting magnitudes
because multiple visits are made to each region.  All surveys
understand the importance of astrometry and all are inserting
astrometric processing into their data processing pipelines.

$\bullet$ {\it Pan-STARRS (Panoramic Survey Telescope And Rapid
Response System):} Led by the Institute of Astronomy of the University
of Hawaii, the first Pan-STARRS telescope (of four planned) achieved
first light in 2006.  Pan-STARRS plans to survey 30000 square degrees
on 0.3\arcsec~pixels to magnitude 24 (single visit).

$\bullet$ {\it SkyMapper:} Based at Siding Spring Observatory,
SkyMapper achieved first light in 2008.  During the planned Southern
Sky Survey, 20000 square degrees will be imaged in six optical filters
on 0.5\arcsec~pixels to magnitude 21 (single visit).  A
shorter-integration survey is planned to sample stars with magnitudes
9--16.

$\bullet$ {\it URAT (USNO Robotic Astrometry Telescope):} Following
the successful USNO CCD Astrograph Catalog (UCAC), URAT (Zacharias
2008) will employ telescopes in both hemispheres to cover the entire
sky in two phases --- phase one for R = 7--18 and phase two to R = 21.
Pixels on the sky are 0.5--0.9\arcsec~in size.  Unique to URAT is its
importance in providing absolute proper motions on the ICRF over the
{\it entire sky}, thereby providing a fundamental reference frame for
both ground- and space-based programs.  A key component is that the
survey bridges the {\it Hipparcos} and {\it Gaia} missions, which
allows proper motions to be distinguished from orbital motions
measured over short periods.

$\bullet$ {\it LSST (Large Synoptic Survey Telescope):} Led by the
LSST Corporation, the magnitude 25 limit (single visit) of LSST is
extraordinary, as are the 0.2\arcsec~pixels used to sample the sky in
six optical filters.  First light is planned for 2015 and several
hundred observations of each piece of sky are likely.  Thus,
parallaxes and proper motions can be determined for huge samples that
will not be observed by any other ground- or space-based effort.

Among the ``lowest hanging fruit'' to be picked from these surveys is
the study of stellar populations.  All stellar types within 10 pc and
most stellar types to 100 pc will have accurate parallaxes measured in
only a few years.  The end of the stellar main sequence falls near
$M_V$ = 20, so all of the above surveys will find all stars within 10
pc.  At 100 pc, the end of the main sequence is $V$ = 25, which will
be sampled only by LSST, but all four surveys will reach at least
M4.0V type, as well as all but the coolest white dwarfs and subdwarfs.
The luminosity function for such stars can be measured once and for
all, and because high quality, multi-color photometric data will also
be collected, credible physical models can be constructed.  Bright
brown dwarfs will also be comprehensively sampled by LSST and
Pan-STARRS.  Even for the reconnaissance of the solar neighborhood, a
synergy of astrometric efforts on the ground and in space is required
because various efforts sample different magnitude ranges --- some
saturate where others are most sensitive, while some reach deeper than
others can reach.  The decade 2010-2020 will, for the first time,
provide that synergy.

%%%%%%%%%%%%%%%%%%%%%%%%%%%%%%%%%%%%%%%%%%%%%%%%%%%%%%%%%%%%%%%%%%%%%%
\vspace{-10pt}
\section{Education in Astrometry}
%%%%%%%%%%%%%%%%%%%%%%%%%%%%%%%%%%%%%%%%%%%%%%%%%%%%%%%%%%%%%%%%%%%%%%

The potential for astrometry to contribute to science is greater than
at any time in history.  A desperate need exists to train young
scientists in astrometry to make the most of this exciting future.  We
have outlined above myriad scientific questions that can be answered
during the next decade by continuing ground-based astrometry, yet all
of these are moot if no one knows the techniques needed to produce the
science.

If we are to take advantage of the science opportunities made possible
by the advent of $\mu$as astrometry, we need to reformulate our study
of reference frames, systems, and the equations of motion in the
context of special and general relativity.  Methods need to be
developed to statistically analyze our data and calibrate our
instruments to levels beyond current standards.  As a consequence, our
curricula must be drastically revised to meet the needs of students in
the 21st century (van Altena \& Stavinschi 2008).  Significant
investments must be made by universities in the US to hire specialists
to teach new generations of scientists how to guide the development,
optimize the use of, and analyze the data from advanced
instrumentation with astrometric capabilities.  This is a critical
problem.  Not one university in the US has hired a specialist to teach
astrometry for many years and there is currently no formal astrometry
course being taught in the US.  Aside from the US Naval Observatory,
not one of the major observatories, e.g., KPNO, CTIO, Gemini, Keck,
Pan-STARRS, LSST, etc.~has a specialist in the field on its staff
(except for D.~Monet, who is doing at least double duty).  Even more
stunning is that the NSF has not funded a single proposal in optical
astrometry for the last two years.  This situation must be rectified
in the coming decade or our national investments in both ground- and
space-based astrometric missions will be severely compromised.

%%%%%%%%%%%%%%%%%%%%%%%%%%%%%%%%%%%%%%%%%%%%%%%%%%%%%%%%%%%%%%%%%%%%%%
\vspace{-10pt}
\section{Astrometry Short List 2010-2020}
%%%%%%%%%%%%%%%%%%%%%%%%%%%%%%%%%%%%%%%%%%%%%%%%%%%%%%%%%%%%%%%%%%%%%%

{\bf 1. We recommend that NSF and NASA consider funding $\sim$10
tenure-track scientists to carry out significant astrometric programs
in university environments where students can learn the tools of the
trade.}  Currently in the United States there are fewer than five
tenured faculty whose primary science is centered around classical
astrometry.  Ideally, this number would be at least doubled during the
next decade to (a) allow the science to advance, (b) provide students
opportunities to learn, and (c) support the complex astrometric needs
for major projects planned that will commence in the next decade and
carry into 2020-2030.

\noindent {\bf 2. We recommend that telescopes with apertures less
than 4m be supported through the next decade so students obtain real
experience making astrometric observations.}  To take advantage of the
upcoming expensive ground- and space-based astrometric efforts,
meaningful, hands-on experience for young astronomers is needed, and
is most easily carried out on relatively small telescopes.  Of course,
appropriately stable imaging instrumentation, ideally with AO for
diffraction limited observations, is also needed.

\noindent {\bf 3. We recommend that in the area of radio astrometry,
upgrades in receivers and data recording equipment be made at the
VLBA.}  Specifically, adding a new receiver capable of observing the
ubiquitous 6.7 GHz methanol masers is needed to map the locations and
motions of hundreds (instead of tens) of star forming regions across
the Milky Way.  Upgrading the VLBA data recording rate by two orders
of magnitude would dramatically improve astrometric accuracy by making
far more background quasars available as positional references.  The
construction of a high frequency ``SKA'' capability would
revolutionize radio astrometry and lead to truly dramatic results
beyond the Local Group.

\noindent {\bf 4. We recommend that a few programs be identified for
multi-decade, consistent, astrometric research so that unique long
time domains can be sampled.}  Perhaps the single most important
reason to support narrow angle, ground-based astrometry during the
next decade is because long-term astrometric series open new parameter
space unmatched by any of the large efforts, which typically have
horizons of five years.  Many small telescopes allow for this
capability, and some large telescopes, e.g., Keck, are making moves
toward long-term possibilities.

\noindent {\bf 5. We recommend that data mining opportunities for
students and database management tools be supported so that the next
generation of astronomers can take advantage of the wealth of data
that will pour out of the astrometric surveys.}  In the age of
terabytes of data/night, we will certainly be limited by our
efficiency in sifting through the data mountains for the few gems that
matter.  Only with astute planning can the exquisite science happen,
or we have forfeited our opportunities and our resources.

%%%%%%%%%%%%%%%%%%%%%%%%%%%%%%%%%%%%%%%%%%%%%%%%%%%%%%%%%%%%%%%%%%%%%%
\vspace{-10pt}
\section{References}
\vspace{-02pt}
%%%%%%%%%%%%%%%%%%%%%%%%%%%%%%%%%%%%%%%%%%%%%%%%%%%%%%%%%%%%%%%%%%%%%%

\scriptsize

\noindent Benedict, G.F., et al. 2008, IAU Symposium 248, A Giant
Step: from Milli- to Micro-arcsecond Astrometry, Eds. Wenjing Jin,
Imants Platais \& Michael A.C. Perryman (Cambridge: CUP) \vskip02pt

\noindent Dahn, C.C., et al. 2002, AJ, 124, 1170 \vskip02pt

\noindent Dinescu, D.I, Girard, T.M., van Altena, W.F., Lopez, C.E.
2005, ApJ 618, L25 \vskip02pt

\noindent Girard, T.M., Korchagin, V.I., Casetti-Dinescu, D.I., van
Altena, W.F., Lopez, C.E., Monet, D.G. 2006, AJ, 132, 1768 \vskip02pt

\noindent Harris, H.C., et al. 2007, AJ, 133, 631 \vskip02pt

\noindent Helmi, A. \& White, S.D.M. 1999, MNRAS 307, 495 \vskip02pt

\noindent Henry, T.J., Jao, W.-C., Subasavage, J.P., Beaulieu, T.D.,
Ianna, P.A., Costa, E., \& M{\'e}ndez, R.A. 2006, AJ, 132, 2360
\vskip02pt

\noindent Honma, M., et al. 2007, PASJ, 59, 889 \vskip02pt

\noindent Jao, W.-C., Mason, B.D., Hartkopf, W.I, Henry, T.J., \&
Ramos, S.N. 2009, arXiv:0902.0555 \vskip02pt

\noindent Luyten, W.J. 1979, LHS Catalogue (Minneapolis: Univ. of
Minnesota Press) \vskip02pt

\noindent Menten, K.M., Reid, M.J., Forbrich, J., \& Brunthaler,
A. 2007, A\&A, 474, 515 \vskip02pt

\noindent Perryman, M.A.C. 2008, IAU Symposium 248, A Giant Step: from
Milli- to Micro-arcsecond Astrometry, Eds. Wenjing Jin, Imants Platais
\& Michael A.C. Perryman (Cambridge: CUP) \vskip02pt

\noindent Platais, I., Kozhurina-Platais, V., Mathieu, R.D., Girard,
T.M., \& van Altena, W.F. 2003, AJ, 126, 2922 \vskip02pt

\noindent Reid, M.J., Menten, K.M., Brunthaler, A., Zheng, X.W.,
Moscadelli, L., \& Xu, Y. 2008, arXiv:0811.0595 \vskip02pt

\noindent Subasavage, J.P., Jao, W.-C., Henry, T.J., Bergeron, P.,
Dufour, P., Ianna, P.A., Costa, E., \& Mendez, R.A. 2009,
arXiv:0902.0627 \vskip02pt

\noindent Thorstensen, J.R., L{\'e}pine, S., \& Shara, M. 2008, AJ,
136, 2107 \vskip02pt

\noindent van Altena, W.F., Lee, J.T., \& Hoffleit, D. 1995, VizieR
Online Data Catalog, 1174, 0 \vskip02pt

\noindent van~Altena, W. \& Stavinschi, M. 2008, IAU Symposium 248, A
Giant Step: from Milli- to Micro-arcsecond Astrometry, Eds. Wenjing
Jin, Imants Platais \& Michael A.C. Perryman (Cambridge: CUP)
\vskip02pt

\noindent van Leeuwen, F. 2007, Hipparcos, the New Reduction of the
Raw Data, Astrophys \& Space Science Lib, Vol. 350, 20, Springer \vskip02pt

\noindent Vieira, K., van Altena, W.F., \& Girard, T.M. 2005,
Astrometry in the Age of the Next Generation of Large Telescopes,
338,130 \vskip02pt

\noindent Vrba, F.J., et al. 2004, AJ, 127, 2948 \vskip02pt

\noindent Zacharias, N. 2008, IAU Symposium, 248, 310

% \vskip10pt
% \noindent {\it Websites}
% \vskip10pt
% 
% \noindent {\it Gaia}:
% http://www.rssd.esa.int/index.php?project=GAIA\&page=index
% \vskip05pt
% \noindent {\it Hipparcos}:
% http://sci.esa.int/science-e/www/area/index.cfm?fareaid=20
% \vskip05pt
% \noindent LSST: http://www.lsst.org/lsst
% \vskip05pt
% \noindent Pan-STARRS: http://pan-starrs.ifa.hawaii.edu/public/
% \vskip05pt
% \noindent RECONS: http://recons.org
% \vskip05pt
% \noindent SkyMapper: http://www.mso.anu.edu.au/skymapper/index.php
% \vskip05pt
% \noindent URAT: http://adsabs.harvard.edu/abs/2009AAS...21347011Z

%%%%%%%%%%%%%%%%%%%%%%%%%%%%%%%%%%%%%%%%%%%%%%%%%%%%%%%%%%%%%%%%%%%%%%
%%%%%%%%%%%%%%%%%%%%%%%%%%%%%%%%%%%%%%%%%%%%%%%%%%%%%%%%%%%%%%%%%%%%%%
%%%%%%%%%%%%%%%%%%%%%%%%%%%%   THE END   %%%%%%%%%%%%%%%%%%%%%%%%%%%%%
\end{document}